\RequirePackage{amsmath}

\documentclass[runningheads]{llncs}

\usepackage{graphicx}
\usepackage{url}
\usepackage{amsmath,bm}
\usepackage{amssymb}
\usepackage{cite}
\usepackage{bbding}
\begin{document}

\title{3D U$^2$-Net: A 3D Universal U-Net for Multi-Domain Medical Image Segmentation\thanks{C. Huang and S. Zhu were supported by Cyrus Tang Foundation \& Zhejiang University Education Foundation. H. Han was supported by the Natural Science Foundation of China (61732004 and 61672496), External Cooperation Program of CAS (GJHZ1843), and Youth Innovation Promotion Association CAS (2018135). This work was done when C. Huang was an intern in MIRACLE.}}
\titlerunning{3D U$^2$-Net: A 3D Universal U-Net for Multi-Domain Segmentation}
%
\author{Chao Huang\inst{1,2}\orcidID{0000-0003-4974-2649} \and
Hu Han\inst{2,3} \and
Qingsong Yao\inst{2} \and
Shankuan Zhu\inst{1(}\Envelope\inst{)} \and
S. Kevin Zhou\inst{2,3(}\Envelope\inst{)}}
%

\authorrunning{C. Huang, et al.}
%
\institute{Chronic Disease Research Institute and Department of Nutrition and Food Hygiene, School of Public Health, and Women's Hospital, School of Medicine, Zhejiang University, Hangzhou 310058, China \\
\email{\{huangchao09,zsk\}@zju.edu.cn} \and
Medical Imaging, Robotics, Analytic Computing Laboratory/Engineering (MIRACLE), Key Lab of Intelligent Information Processing of Chinese Academy of Sciences (CAS), Institute of Computing Technology, CAS, Beijing 100190, China\\
\email{\{hanhu,zhoushaohua\}@ict.ac.cn} \and
Peng Cheng Laboratory, Shenzhen, China}

\maketitle              
\begin{abstract}
Fully convolutional neural networks like U-Net have been the state-of-the-art methods in medical image segmentation. Practically, a network is highly specialized and trained separately for each segmentation task. Instead of a collection of multiple models, it is highly desirable to learn a universal data representation for different tasks, ideally a single model with the addition of a minimal number of parameters steered to each task. Inspired by the recent success of multi-domain learning in image classification, for the first time we explore a promising universal architecture that handles multiple medical segmentation tasks and is extendable for new tasks, regardless of different organs and imaging modalities. Our 3D Universal U-Net (3D U$^2$-Net) is built upon separable convolution, assuming that {\it images from different domains have domain-specific spatial correlations which can be probed with channel-wise convolution while also share cross-channel correlations which can be modeled with pointwise convolution}. We evaluate the 3D U$^2$-Net on five organ segmentation datasets. Experimental results show that this universal network is capable of competing with traditional models in terms of segmentation accuracy, while requiring only about $1\%$ of the parameters. Additionally, we observe that the architecture can be easily and effectively adapted to a new domain without sacrificing performance in the domains used to learn the shared parameterization of the universal network. We put the code of 3D U$^2$-Net into public domain. \footnote{\url{https://github.com/huangmozhilv/u2net_torch/}}
\end{abstract}
\section{Introduction}
Image segmentation is crucial for clinical practice and health research. Fully convolutional neural networks (CNNs) like U-Net~\cite{ronneberger2015u} have been the dominant approach in automatic medical imaging segmentation\cite{cciccek20163dunet, milletari2016vnet}. A practical segmentation model is learned by customizing a neural network architecture for a certain task or dataset and training it from scratch \cite{milletari2016vnet, roth2015deeporgan, savioli2018v}. \cite{karani2018lifelong} learned a single segmentation CNN for brain datasets acquired with different scanners and/or protocols. Notwithstanding being powerful, these models are difficult to extend to new tasks with unseen contents because of the highly specialized design. \cite{isensee2018nnunet} took one step further by presenting a self-adapting framework for various tasks, yielding mutually independent models for each task.
On the contrary, human experts can easily learn to tackle multiple tasks and generalize to new tasks on the basis of acquired skills. Multiple previous works explored multi-task segmentation, wherein all organs of interest appear in the same image~\cite{lay2013rapid,roth2017hierarchical}. Here we consider a more realistic and challenging scenario: for a given dataset, only a local region of the human body is scanned and only one or several anatomical structures within the image are annotated. \cite{moeskops2016deep} focused on a similar topic and trained one single CNN on three tasks, however, the trained model was designed as such that it cannot be extended to other tasks. From this point of view, an effective and efficient method for image segmentation remains an open problem.

Bilen et al. \cite{bilen2017universal, rebuffi2017learning, rebuffi2018efficient} suggested that there might exist a universal data representation across different visual domains. Specifically, they introduced a new competition called Visual Decathlon Challenge \footnote{\url{https://www.robots.ox.ac.uk/~vgg/decathlon/}}, aiming to simultaneously model ten visual domains of different styles and contents, e.g., internet images, hand-written characters, sketches, planktons, etc. \cite{rebuffi2017learning}. They referred to such a new topic as ``multi-domain learning" and realized the universal representation by piggybacking parallel residual adapters on the model pre-trained with ImageNet. However, their work exclusively focuses on image classification. Naturally, one question occurs to us: {\it is it possible to build a single neural network that can deal with medical segmentation tasks from different domains?}

To achieve this goal, we draw inspiration from previous studies \cite{chollet2017xception, guo2019depthwise}, particularly \cite{guo2019depthwise} which won the first place in the Visual Decathlon Challenge to date. \cite{guo2019depthwise} believed that \cite{rebuffi2018efficient} ignored the structural heterogeneity of various domains and attempted to address the issue by leveraging depthwise separable convolution. While standard convolution conducts the spatial and channel-wise computation at once, such convolution factors the computation into two sequential steps: first, depthwise convolution applies an independent convolutional filter per input channel, and then a pointwise convolution follows to linearly combine the output across all channels for every spatial location. The basic building block of their multi-domain network comprises a cohort of parallel channel-wise convolutions, one per domain, followed by one pointwise convolution shared by all domains. The insight is that the former is better to capture domain-specific spatial patterns while the latter probes the sharable cross-channel interdependencies. In this paper, we claim to note ``depthwise separable convolution" as ``separable convolution" and ``depthwise convolution" as ``channel-wise convolution" to avoid confusion with the depth dimension of the image volume.

Based on the separable convolution as introduced above, our work proposes a universal architecture for multi-domain medical image segmentation. The main idea behind is rather intuitive yet powerful: a basic network is first designed on the ground of 3D U-Net\cite{ronneberger2015u,cciccek20163dunet} (or  V-Net\cite{milletari2016vnet}), and then any $3 \times 3 \times 3$ standard convolution with a stride of 1 is substituted by separable convolution similar to \cite{guo2019depthwise}. However, our approach substantially differ from \cite{guo2019depthwise} as following: (1) their work focuses on image classification which is fundamentally different from image segmentation here. (2) they obtain the ultimate multi-domain architecture in three steps: First, pre-training a ResNet-26 modified with separable convolution on ImageNet; Second, freezing and transferring the pointwise convolution weights to new network; Thirdly, training the new network on each domain separately and stacking the channel-wise convolutions together while sharing the pointwise convolution weights from the pre-trained model. Nevertheless, we manage to train across the domains together to obtain the final model. (3) we further adapt our universal network to a new domain by simply adding new channel-wise convolutions. To the best of our knowledge, this is the first time to learn an extendable universal network for multi-domain medical image segmentation.

\section{Methods}
\subsection{Problem definition}
Let $\{D_1,D_2,\cdots,D_T\}$ be a set of $T$ image domains, among which domain $D_t$ consists of two paired image spaces of $\{X_t,Y_t\}$. $X_t\in \mathbb{R}^{C_t\times D\times H \times W}$ is the input image space and $Y_t\in \mathbb{R}^{C_t^\prime \times D \times H \times W}$ is the output image space, i.e., segmentation masks. $D$, $H$ and $W$ are the spatial depth, height and width. $C_t$ and $C_t^\prime$ are the numbers of imaging modalities and segmentation classes specific to each domain. To work well on all domains, our universal network contains domain-specific parameters as well as shared parameters. Let $\theta_t$ be the domain-specific parameters for domain $D_t$ and $\theta_u$ be the universally shared parameters by all domains. Assuming $\{x_{t,i}, y_{t,i}\}$ as the $i^{th}$ training pair of domain $D_t$, then the output $\hat Y$ of the neural network $F(X)$ is
\begin{align}
& \hat y_{t,i} = F(x_{t,i};\theta_u,\theta_t). \label{eq:output1}
\end{align}

\begin{figure}[t]
\centering
\includegraphics[width=0.8\textwidth]{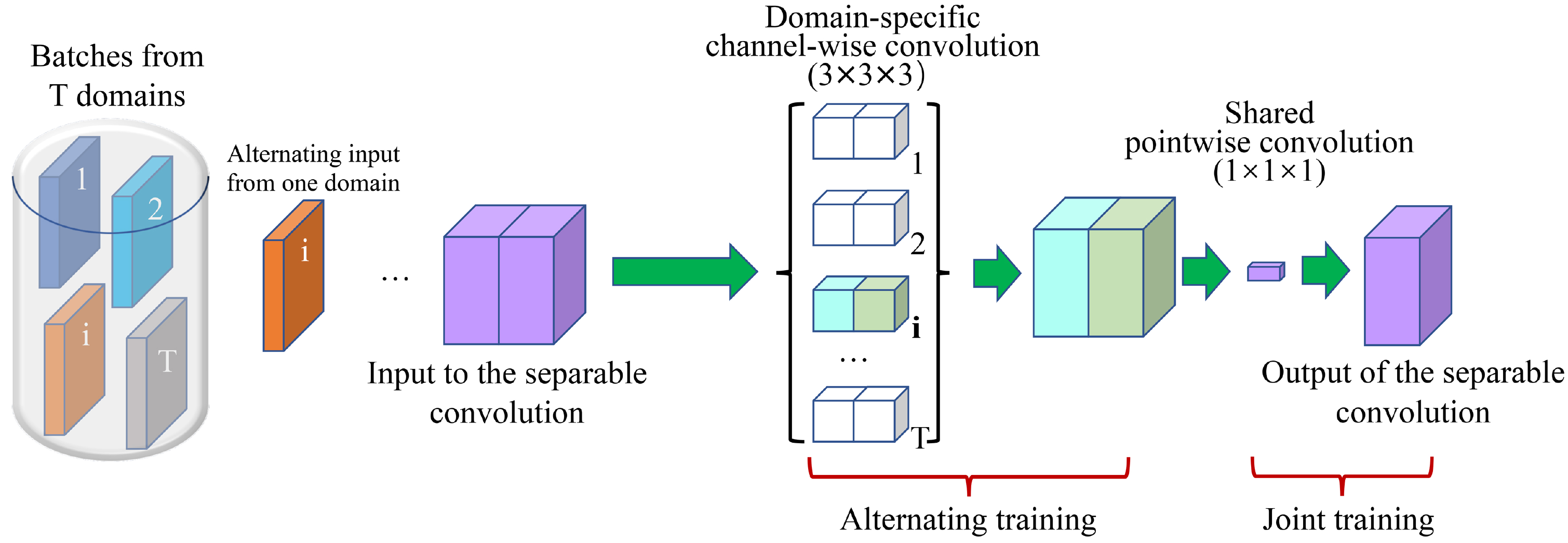}
\caption{Domain adapter based on separable convolution.} \label{fig_depth_sep}
\end{figure}

\subsection{Domain adapter}
Domain adapter, the key component to ensure the success of our universal network, consists of both domain-specific parameters and shared parameters and is built upon separable convolution in place of standard convolution. 

In standard convolution with filter $W \in \mathbb{R}^{3\times 3\times 3\times C \times C^\prime}$ applied to an input tensor $U \in \mathbb{R}^{C\times D\times H \times W}$, the output tensor $\hat U \in \mathbb{R}^{C^\prime \times D\times H \times W}$ is obtained by applying $C^\prime$ filters $w \in \mathbb{R}^{3\times 3\times 3\times C}$ on the input in parallel and concatenating the $C^\prime$ output feature maps. A simple calculation tells that the total number of filter parameters in the above filters is $27 * C * C'$. Also, when training the models for the $T$ domains separately, the number of parameters grows $T$ times! 

In separable convolution, the computation is factorized into two sequential steps. The first step applies $C$ channel-wise filters $w \in \mathbb{R}^{3\times 3\times 3}$ to each channel of the input in parallel and concatenate the $C$ output feature maps together. Here, {\it each domain has its own channel-wise filters}.  The second step then applies $C^\prime$ pointwise filters $w \in \mathbb{R}^{1\times 1\times 1 \times C}$ to output the final feature maps of $C^\prime$ channels. Here, {\it all domains share the same pointwise filters}. A simple calculation tells that the total number of weights in the above filters is $27*C*T+C*C'$. How to assemble the domain-specific channel-wise convolutions and the shared pointwise convolution to form a domain adapter is illustrated in Fig.~\ref{fig_depth_sep}. 

\begin{figure}[t]
\includegraphics[width=\textwidth]{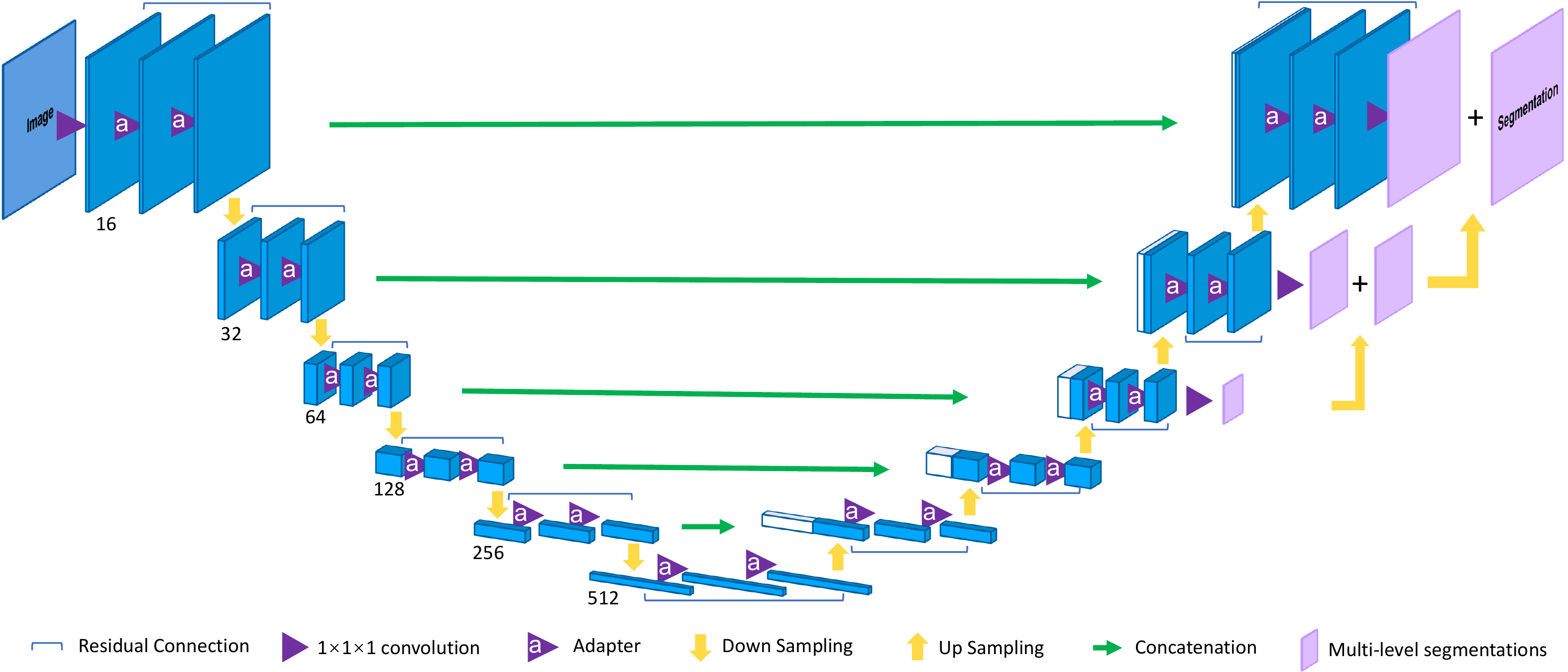}
\caption{The proposed 3D Universal U-Net (3D U$^2$-Net).} \label{fig_network}
\end{figure}

\subsection{3D Universal U-Net (3D U$^2$-Net)}
As shown in Fig.~\ref{fig_network}, our universal network architecture is based on a basic network with six components: (1) input; (2) encoder path; (3) bottleneck block; (4) decoder path; (5) deep supervision branch; and (6) output. Channels of the input and output could vary according to the number of imaging modalities and classes of different domains. In general, the input layer uses 16 filters. The encoder and decoder paths both contain five levels at different resolutions. Residual connection is applied within each level. Skip connection is employed to preserve more contextual information from the encoder counterpart for decoder path\cite{ronneberger2015u}. Inspired by \cite{kayalibay2017cnn}, we incorporate a deep supervision branch alongside the end of decoder path via element-wise sum of multi-level segmentation maps to boost the final localization performance.
To construct the universal network, domain adapters detailed above are inserted into basic network to replace any standard $3 \times 3 \times 3$ convolution with a stride of 1. 

\subsection{Loss function}
 A hybrid loss function is employed by combining Lovász-Softmax loss \cite{berman2018lovasz}, capable of improving intersection-over-union segmentation scores, and focal loss \cite{lin2017focal}, aimed to alleviate class imbalance. During training the universal model, we sample a batch from each dataset in a round-robin fashion, allowing each domain to contribute to the shared parameters. Assuming that for the $n$th iteration the batch data pair $\{x_t, y_t\}$ is from domain $D_t$, the corresponding loss $L_n$ is
\begin{align}
& L_n = L_L(x_t,y_t; \theta_u, \theta_t)  + L_f(x_t,y_t; \theta_u, \theta_t), \label{eq:loss_n}
\end{align}
where $\theta_t$ be the domain-specific parameters for domain $D_t$ and $\theta_u$ be the universally shared parameters of the neural network. $L_L$ is the Lovász-Softmax loss and $L_f$ is the focal loss counterpart.

\section{Experimental Results}
In this section, we present extensive experiments to evaluate the proposed 3D U$^2$-Net in dealing with medical multi-organ segmentation: (1) independent models, aimed to reproduce the traditional methods, are obtained by training the basic network for each base domain separately; (2) shared model, which aims at investigating whether all parameters of a model can be shared by all domains and thus is gained by training the single basic network with all base domains together; and (3) universal model, which is our ultimate goal and is achieved by training the universal architecture with all base domains simultaneously. Notably, the first two represent two extreme multi-organ segmentation approaches and serve as baselines for the universal model. Additionally, we test the generalizability of both the shared model and universal model on one new domain.

\subsubsection{Datasets:} We use six public datasets from the Medical Segmentation Decathlon challenge\footnote{\url{https://decathlon.grand-challenge.org/}} as introduced by \cite{simpson2019large}. The first five datasets are considered as base domains and are used to train the universal model. On the other hand, the last dataset is treated as the new domain and is used to test the adaptiveness of the universal model. Basic characteristics of the datasets are shown (Table~\ref{tab_datasets}). For each dataset, 80\% of the samples are randomly extracted for training, while the remaining 20\% are used as testing data.

\begin{table}[t]
\caption{Basic characteristics of the datasets.}\label{tab_datasets}
\tiny
\begin{tabular}{|l|p{1cm}|l|l|l|}
\hline
Task & Modality & Data size & Image shape & Voxel spacing\\ 
\hline
Base01\_Heart & MRI & 20 & (90$\sim$130)$\times$320$\times$320 & 1.37$\times$1.25$\times$1.25\\
\hline
Base02\_Liver & CT & 131 & (74$\sim$987)$\times$512$\times$512 & (0.7$\sim$5)$\times$(0.557$\sim$1)$\times$(0.557$\sim$1)\\
\hline
Base03\_Hippocampus & MRI & 260 & (24$\sim$47)$\times$(40$\sim$59)$\times$(31$\sim$43) & 1$\times$1$\times$1\\
\hline
Base04\_Prostate & T2, ADC & 32 & (11$\sim$24)$\times$(256$\sim$384)$\times$(256$\sim$384) & (3$\sim$4)$\times$(0.6$\sim$0.75)$\times$(0.6$\sim$0.75)\\
\hline
Base05\_Pancreas & CT & 281 & (37$\sim$751)$\times$512$\times$512 & (0.7$\sim$7.5)$\times$(0.605$\sim$0.977)$\times$(0.605$\sim$0.977)\\
\hline
New\_Spleen & CT & 41 & (31$\sim$168)$\times$512$\times$512 & (1.25$\sim$7.5)$\times$(0.535$\sim$0.977)$\times$(0.535$\sim$0.977)\\
\hline
\end{tabular}
\end{table}

\vspace{-10pt}
\subsubsection{Preprocessing:} The datasets are highly diverse in terms of modality, image size and voxel spacing. Pre-processing procedures are conducted as below: (1) all images are cropped to the region of nonzero values, thereby reducing the image size to alleviate computation burden; (2) all images are resampled to the median voxel spacing of the corresponding dataset to retain spatial semantics; (3) for each patient, the image is clipped to the [2.0,98.0] percentiles of the intensity values of the entire image, followed by Z-score normalization with the mean and standard deviation of the image for each modality; and (4) the following data augmentation are applied: random elastic deformation, random rotation, random scaling and random mirroring. Data augmentation is done ``on-the-fly" during training with batch generators\footnote{\url{https://github.com/MIC-DKFZ/batchgenerators/}}, a python package maintained by the Division of Medical Image Computing at the German Cancer Research Center.

To accommodate the limited GPU memory, we train the network with patches randomly sampled from the whole images. While for inference, the patches are generated with a sliding window moving across the entire image with a stride of half patch size. As for the shared model and universal model, the input batch is of two patches with a size of $128\times128\times128$ and the number of down-sampling operations is set to 6. However, for the independent models, we adjust the input patch size and the resolution levels for each domain considering the image size in order to maximize the utilization of computation resources. If the median shape is smaller than $128\times128\times128$, we toggle between the input patch size and batch size to have the patch size of the same aspect ratio as the median shape. The number of down-sampling operations per axis is set until the feature map size of the deepest layer reaches as small as 8. Specifically, to prepare the patches for shared model and universal model, we first extract a patch of size as in the independent model and then resize it to the above target patch size.

\vspace{-10pt}
\subsubsection{Implementation details:} The network is implemented in Pytorch 1.0.1 on an NVIDIA V100 GPU. The ADAM optimizer is applied with an initial learning rate of $3\times10^{-4}$ and a weight decay of $10^{-5}$. An epoch is defined as an iteration over 250 batches. Exponential moving average, $\textit{l}^{t}_{MA}$, is monitored for training loss for every 30 epochs. The learning rate is reduced by a factor of 5 as long as $\textit{l}^{t}_{MA}$ does not decrease by $5\times10^{-4}$. We terminate the training once the learning rate is below $10^{-8}$. During training the shared and universal models, we apply a round-robin fashion to feed the network sample batches from each domain in turn, so as to allow all the domains to contribute to the final model equally. The results are presented on the testing data.

\vspace{-10pt}
\subsubsection{Quantitative results of base domains:} Table~\ref{res_base_domains} lists the mean Dice scores of the three models on each base domain. Comparing along the columns, we observe that the independent models obtain the highest scores on most domains and yield the highest overall mean score. However, strikingly both the shared model and the universal model achieve moderate performance for most domains comparable to the independent models, and gains significant increase regarding to peripheral zone (PZ) and transition zone (TZ) of Base04\_Prostate. Compared to the shared model, we further observe that the universal model is better in the segmentation of pancreas for Base05\_Pancreas. Besides, the universal model gets an overall higher mean score across all domains in comparison to the shared model. The increase in overall performance could be attributed to the use of domain-specific parameters that can agree with each domain well.

\begin{table}[t]
\caption{Quantitative results on base domains.}\label{res_base_domains}
\centering
\begin{tabular}{|c|c|c|c|c|c|c|c|c|}
\hline
 \multicolumn{1}{|c|}{}& Base01\_ & Base02\_ &  \multicolumn{2}{|c|}{Base03\_} & \multicolumn{2}{|c|}{Base04\_} & {Base05\_} & {} \\
 \multicolumn{1}{|c|}{}& Heart & Liver &  \multicolumn{2}{|c|}{Hippocampus} & \multicolumn{2}{|c|}{Prostate} & Pancreas & {} \\
\hline
 \multicolumn{1}{|c|}{(Dice\%) } & Left\_atrium & Liver & Anterior & Posterior & PZ & TZ & Pancreas & Mean\\
\hline
 \multicolumn{1}{|c|}{independent} & 93.26 & 95.02 & 89.62 & 87.74 & 58.39 & 87.18 & 78.78 & 84.28 \\
\hline
 \multicolumn{1}{|c|}{shared} & 92.73 & 93.40 & 89.25 & 87.30 & 68.38 & 89.30 & 57.57 & 82.56 \\
\hline
 \multicolumn{1}{|c|}{universal} & 91.98 & 93.54 & 89.34 & 87.05 & 68.50 & 89.21 & 62.08 & 83.10\\
\hline
\end{tabular}
\vspace*{-5pt}
\end{table}

\vspace{-10pt}
\subsubsection{Model complexity:} When investigating the complexity of the models, we exclude the input layer, last layer and deep supervision branch as they are never shared across domains. The basic network used in the shared model is considered as reference. The number of parameters are computed and displayed in Table~\ref{model_complexity}(a). Obviously the proposed 3D U$^2$-Net 
requires the least parameters, indicating that it can perform effectively across various domains. The overall number of parameters from the universal model is around {\bf 1\%} of that of all independent models, while the two obtain comparable segmentation accuracy.

\vspace{-10pt}
\subsubsection{Quantitative results of a new domain:} Furthermore, we conduct experiments to illustrate the effectiveness of adapting the trained shared model or universal model to a new task, which are implemented by freezing the corresponding shared pointwise convolutions or standard convolutions and adding and training all other domain-specific modules like input layer and channel-wise convolutions in parallel to the structures of the same kindred for this domain.
Table~\ref{model_complexity}(b) shows that the universal model performs better for the new domain `New\_Spleen' in comparison to the shared model, therefore indicating a superior generalization ability over the latter. This adds further evidence of the effectiveness of the domain-specific parameters. The universal model is adaptive to new domain with a few extra parameters, i.e., 0.3\% compared to the traditional independent model, which is exactly what we anticipate in this paper.

\begin{table}[t]
\caption{(a) Model complexity. (b) Quantitative results on a new spleen domain.}\label{model_complexity}
\centering
\begin{tabular}{|c|c|c|c|c|c|c|c|c|}
\hline
 \multicolumn{1}{|c|}{}& (a) \#Par & (a) Ratio && (b) New\_Spleen  -- Dice\%& (b) \#Added Par\\
\hline
 \multicolumn{1}{|c|}{independent} & 126.7M & $4.1 \times$ && 92.37 & 30.7M\\
\hline
 \multicolumn{1}{|c|}{shared} & 30.7M & $1\times$ && 90.67 & 0\\
\hline
 \multicolumn{1}{|c|}{universal} & 1.7M & $0.06\times$ && 91.60 & 0.1M\\
\hline
\end{tabular}\vspace*{-5pt}
\end{table}

\section{Conclusions}
In summary, we present a novel universal neural network named 3D U$^2$-Net for multi-organ segmentation problem, filling the gap of extendable multi-domain learning in image segmentation. Experimental results demonstrate that the proposed approach, with only a tiny portion of the parameters, obtains the segmentation performance comparable to the independent models trained in the traditional manner. As CT and MRI images are routine images on hand and the amount of human organs is constant, the universal model for multi-organ segmentation can be fully developed soon in the near future. Besides, the proposed framework could extend to many other multi-domain applications and thus facilitate the translation of neural networks to clinical practice.

%
%
%
\bibliographystyle{splncs04}
\bibliography{miccai2019}
%
\end{document}